\documentclass[12pt]{iopart}
\usepackage{epsfig}

\begin{document}

\article{}{Semiclassical description of the kinematically complete
experiments}

\author{F J\' arai-Szab\' o\dag\ and L Nagy\dag}

\address{\dag\  Faculty of Physics, Babe\c{s}-Bolyai University, str. 
Kog\u{a}lniceanu 1, RO-400084 Cluj-Napoca, Romania}

\begin{abstract}
Based on the semiclassical, impact parameter method a theoretical model is 
constructed to calculate fully differential cross sections for single 
ionization of helium by impact with fast C$^{6+}$ ions. Good agreement with the
 experiment is achieved in the scattering plane, while in the perpendicular 
plane a similar structure to that observed experimentally is obtained. The 
contribution of different partial waves to the cross section is also 
investigated.
\end{abstract}
\ead{lnagy@phys.ubbcluj.ro}  

The most complete information about ionization processes in atomic collisions 
is provided by fully differential cross sections. These quantities describe the 
entire energy and angular distribution of the ionized electron, residual ion 
and projectile.

Recently, interesting data for the complete electron emission pattern in single 
ionization of helium by the impact of C$^{6+}$ ions for certain momentum 
transfers have been reported \cite{Schulz2003a,Schulz2004}. The three 
dimensional images were generated using experimentally measured fully 
differential cross section values. These experiments were performed on a 
cold-target-recoil-ion-momentum spectrometer (COLTRIMS) apparatus. The results 
show the characteristic double-lobe structure with a binary peak and a smaller 
recoil peak.

Several theoretical calculations exist 
\cite{Madison2003,Foster2004a,Foster2004b} which are able to reproduce the 
experimental data in the scattering plane (determined by the momentum of the 
scattered projectile and the momentum transfer vectors). Right after the 
publication of the first experimental results of fully differential cross 
section measurements for ionization by fast ion impact an intense debate 
existed concerning the discrepancy between experiment  and theoretical 
calculations (mainly performed using the CDW-EIS method) in the plane 
perpendicular to the momentum transfer. Here, the theoretical results are 
essentially isotropic and do not show the observed peak structures 
perpendicular to the beam direction. Some authors have suggested that it may be
important to include into calculations the internuclear interaction \cite{Ciappina2006}. On 
the other side, very recently, the importance of taking into account the 
uncertainties of the experimental measurements and to perform a convolution of 
the theoretical results on the experimental resolution \cite{Fiol2006} was 
proved. At the same time it was suggested that all aspects of the experimental 
resolution may be included into theories by the use of a quantum-theory-based 
Monte Carlo event generator \cite{Durr2007}. In this paper the authors 
conclude that the structures observed in the perpendicular plane may be 
explained only partly by the experimental uncertainties.

In the present work a theoretical model is constructed to calculate fully 
differential cross sections for single ionization of helium by the impact of 
fast C$^{6+}$ ions. The constructed model is based on the first order, 
semiclassical, impact parameter approximation. The aim of this work is to 
explore, how the semiclassical, impact parameter approximation may be used to 
calculate fully differential cross sections. The main problem is to assign a 
value of the impact parameter for a given momentum transfer and electron energy 
and ejection angle. A partial wave analysis for different ejection direction is 
also performed.

In order to study the ionization process of helium produced by fast charged 
projectiles, first the ionization amplitudes have to be calculated. In the
semiclassical approximation the projectile is treated separately and it moves
along a classical trajectory. This implies that 
only the electronic system needs to be described by a time-dependent 
Schr\"odinger equation, while the projectile follows the classical laws of 
motion. Using the first-order perturbation theory, the transition amplitude may 
be written as
\begin{equation}
a^{(1)} = -i \int_{-\infty}^{+\infty}{dt\,e^{i(E_f - E_i)t} \langle f | V_1(t) 
+ V_2(t) | i \rangle }\,,
\end{equation}
where $i$ and $f$ represent the initial and final electronic states of the 
target system, respectively. $E_i$ and $E_f$ are the energies of the 
corresponding (unperturbed) states of the system while $V_1(t)$ and $V_2(t)$ 
denote time-dependent interactions between the projectile and the electrons.

The initial state of the dielectronic system is described by a Hartree-Fock 
wavefunction \cite{Clementi1974}, while the final state is described 
by a symmetric combination of a hydrogenic and a continuum wavefunction
\begin{eqnarray}
| i\rangle &=& | i^{(1)}_b \rangle | i^{(2)}_b \rangle \nonumber \\
| f\rangle &=& \frac{1}{\sqrt{2}} \left(| f_b^{(1)} \rangle | f_c^{(2)} \rangle 
+ | f_c^{(1)}\rangle | f_b^{(2)} \rangle\right)\,.
\end{eqnarray}
Here indexes $b$ and $c$ represent the bound and continuum states, 
respectively, while the indexes $(1)$ and $(2)$ are the labels of electrons. 
The continuum wavefunction is calculated in the mean field of the final He$^+$ 
ion.

With the use of the above described wavefunctions, the ionization probability 
amplitude depending on the momentum transfer vector, ejected electron energy 
and ejection angles is reduced to a one-electron amplitude
\begin{equation}
a^{(1)} = -\frac{i\sqrt{2}}{v} \langle f_b | i_b \rangle 
\int_{-\infty}^{+\infty}{dz\,e^{i \frac{E_f - E_i}{v}z} \langle f_c | V_1(t) | 
i_b \rangle}\,.
\end{equation}
This amplitude is calculated expanding the final continuum-state wavefunctions 
into partial waves. In this way amplitudes for transitions to ionized states 
with different angular momenta ($a_{l_fm_f}^{(1)}$) are obtained.

The fully differential cross sections relative to the momentum transfer value, 
ejected electron energy and electron ejection angles are obtained by the 
relation
\begin{equation}
\frac{d^5\sigma}{dE\;d\theta\;d\phi\;dq\;d\phi_q} = B\;\left| \sum_{l_f,m_f} 
a_{l_fm_f}^{(1)} ({\mathbf B}) \right| ^2 \left|
\frac{dB}{dq}\right|,\label{fdcs1}
\end{equation}
where ${\mathbf B}$ is the impact parameter vector and $l_f$ and $m_f$ are 
quantum numbers of the partial waves describing the ejected electron.

In order to assign an impact parameter to a momentum transfer as a first
approach, the projectile deviation angle is calculated 
using the Rutherford scattering formula, as if we would have only elastic 
scattering. Assuming that the momentum transfer modifies only the direction of 
the projectile's momentum vector and applying some approximations valid for 
small projectile deviations, the impact parameter corresponding to a certain 
momentum transfer will be
\begin{equation}
B = \frac{2 Z_{\rm proj}\;Z_{\rm targ}}{v_p q}\,,
\label{Bq}
\end{equation}
where $Z_{\rm proj}$ is the charge of the projectile, $Z_{\rm targ}$ is the 
effective charge of the target seen by the projectile, $v_p$ is the projectile 
velocity and $q$ denotes the momentum transfer. This means that to a certain 
value of the momentum transfer is assigned a value of the impact parameter 
regardless to the ejection angle of the electron.

We have applied the above outlined model for the ionization of helium induced 
by 100 MeV/u C$^{6+}$ projectiles.

\begin{figure}
\begin{center}
\epsfxsize=4.5cm
\epsfbox{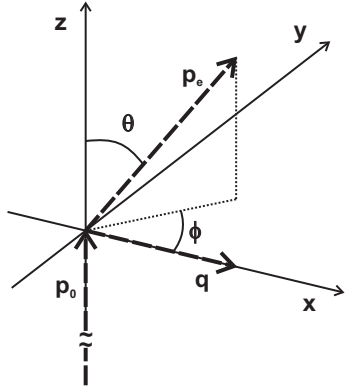}
\end{center}
\caption{\label{coordinates}Sketch of the used coordinate system.}
\end{figure}

In these calculations the used coordinate system is sketched in figure 
\ref{coordinates}. The initial projectile direction is along the $z$ axis and 
the momentum transfer vector $\mathbf q$ is pointing nearly in $x$ direction. 
Standard spherical coordinates are used with azimuthal angle $\theta$ measured 
relative to the projectile beam direction and with polar angle $\phi$ measured 
in the $xy$ plane relative to the $x$ axis.

We calculate fully differential ionization cross sections for an ejected 
electron energy of $E_e = 6.5$ eV and a momentum transfer of $q = 0.75$ a.u. 
Calculating the impact parameter value with expression (\ref{Bq}) we get 
$B=0.253 \div 0.506$, depending on the effective value of target's charge 
($Z_{\rm target} = 1 \div 2$).

\begin{figure}
\begin{center}
\epsfxsize=8cm
\epsfbox{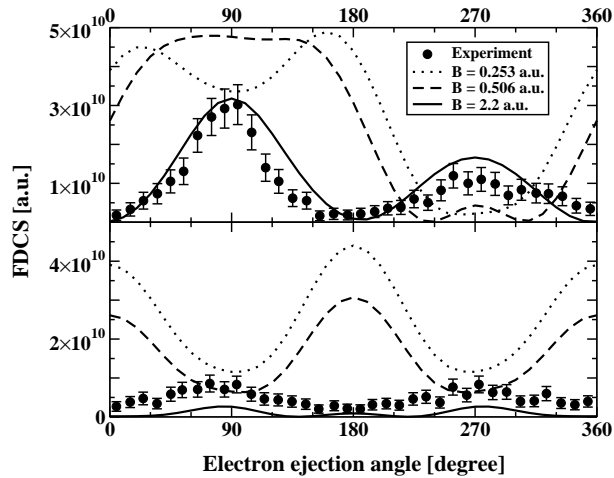}
\end{center}
\caption{\label{fixedip}Fully differential cross sections in the scattering
(top) and perpendicular (bottom) planes, calculated using the Rutherford-type
model using different impact parameters in comparison with experiments
\cite{Fiol2006} for ionization of helium by 100 
MeV/u C$^{6+}$ projectile. The ejected electron energy is $E_e = 6.5$ eV and 
the momentum transfer is $q = 0.75$ a.u.}
\end{figure}

Figure \ref{fixedip} shows theoretical cross section values in two different 
cuts from the 3D theoretical data. The curves show single ionization cross 
section values as a function of electron ejection angle $\theta$. The top panel 
shows the scattering plane characterized by $\phi = 0 \;{\rm or}\; \pi$. The 
bottom panel shows the plane perpendicular to the momentum transfer with $\phi 
= \pi/2 \;{\rm or}\; 3\pi/2$. From figure \ref{fixedip} is immediately 
observable, that the curves representing impact parameter values in the range 
$B=0.253 \div 0.506$ are in disagreement with the experimental results in both 
planes. This means that  the simple model describing the projectile motion as a 
simple Rutherford scattering is not a quite valid description.

Other calculations with higher impact parameter values have also been 
performed. One of these results corresponding to an impact parameter of 2.2 
a.u. is drawn with solid line. In the scattering plane one can observe the 
presence of the characteristic double-lobe structure with the binary peak at 
$\theta = 90^o$ and the recoil peak at $\theta = 270^o$, in agreement with 
experiments. The agreement is worse in case of recoil peak having larger 
theoretical cross section values than the experimental ones. This difference in 
the recoil peak region becomes more accentuated in case of larger momentum 
transfers.

Our attempt to use the simplest Rutherford formula to describe the projectile 
scattering and obtain a correct impact parameter failed. In contrast with the 
elastic scattering, there is no direct correspondence between the momentum 
transfer and impact parameter \cite{Ullrich1997}, the impact parameter depends 
also on the ejected electron energy and angle.

We have investigated several models for obtaining the impact parameter. Results 
in good agreement with the experimental data have been obtained for those, 
which suppose larger impact parameter for the binary peak (where most of the 
momentum transfer is taken by the electron) and smaller impact parameter for 
the recoil peak (where most of the momentum transfer is taken by the target 
nucleus).

A simple calculation is sketched in this sense by the use of the transverse 
momentum balance \cite{Ullrich1997}, meaning that the momentum transfer $q$ is 
the sum of the transverse components of the momenta of the electron and the 
residual ion. This vectorial relation may be written in scalar form as
\begin{equation}
p_{T\perp}^2 = p_{e\perp}^2 + q^2 - 2p_{e\perp}q\cos\phi\,,
\end{equation}
where $p_{T\perp}$ is the transverse momentum taken by the residual ion and 
$p_{e\perp}$ is the transverse momentum of the ionized electron ($p_{e\perp} = 
p_e \sin \theta $). Further we assume, that the impact parameter is related to 
the momentum transfer to the residual ion, and take into account the 
projectile-electron interaction separately. In these conditions the impact 
parameter is obtained to be
\begin{equation}
B = \frac{2 Z_{\rm proj}\;Z_{\rm targ}}{v_p \sqrt{p_{e\perp}^2 + q^2 - 
2p_{e\perp}q\cos\phi}}\,.
\end{equation}
In case of binary peak, one has to deal with $\phi = 0$ while in case of recoil 
peak the value of the angle $\phi$ is 180$^o$. This means that higher impact 
parameters have to be used in case of binary peak than for recoil peak. The 
numerical calculations show higher impact parameter values than the previously 
investigated simple case. In case of an ejected electron energy of $E_e = 6.5$ 
eV and a momentum transfer of $q = 0.75$ a.u. using an effective charge of 
$Z_{\rm target} = 1$ impact parameters of 4.3 and 0.176 a.u. may be obtained 
for binary and recoil peak, respectively.

The two sketched possibilities are two extreme descriptions. The 
Rutherford model treats the residual ion and the electron as one system on 
which the projectile is scattered. The second model treats separately the 
electron and the residual ion. However, the reality may stand between these 
descriptions, while prior the ionization process the target is one single 
system and after the ionization the ionized electron and the recoil ion 
interact separately with the projectile.

In order to find the correct combination of the two extremes and to determine 
the impact parameter values for the binary and recoil peak region, an empirical 
method is used. Cross section values for binary and recoil peaks are considered 
as a function of the impact parameter value and the best impact parameter 
values for binary and the recoil peak region are selected based on the 
experimental data available in the scattering plane. The transition between these 
impact parameter values is realized smoothly in the $0<\theta<50^o$ and 
$130^o<\theta<180^o$, $90^o<\phi<270^o$ transition regions.

In case of electron ejection energy $E_e = 6.5$ eV and momentum transfer 
$q=0.75$ a.u. the experimental results in the scattering plane show cross sections 
of $3\times 10^{10}$ a.u and $1.1 \times 10^{10}$ a.u. for binary peak and 
recoil peak, respectively. In order to obtain these values,
impact parameters of 2.2 and 0.7 a.u are chosen for the binary peak and for the
recoil peak regions, respectively.

\begin{figure}
\begin{center}
\epsfxsize=8cm
\epsfbox{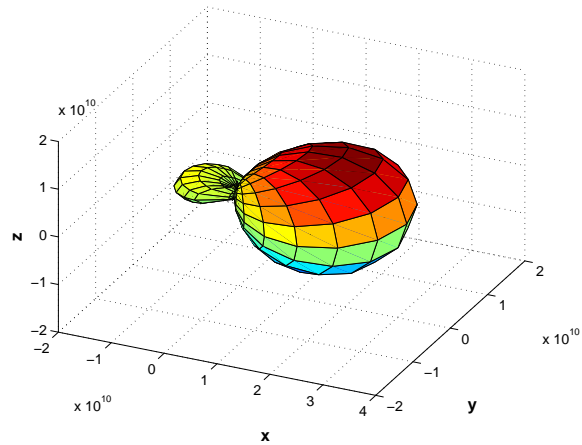}
\end{center}
\caption{\label{3d1}Theoretically obtained 3D image of the electron emission 
pattern for single ionization of helium produced by 100 MeV/u C$^{6+}$ 
projectile impact. The ejected electron energy is $E_e = 6.5$ eV and the 
momentum transfer $q=0.75$ a.u.}
\end{figure}

Figure \ref{3d1} shows the theoretically obtained 3D image of the electron 
emission pattern for the studied case. The results are obtained using the 
previously determined impact parameter values. On the image one can observe the 
presence of the characteristic double-lobe structure towards the x axis with 
binary peak at $\theta = 90^o$, $\phi = 0^o$ and recoil peak at $\theta = 
90^o$, $\phi = 180^o$. And now, by using this semi-empirical model, the 
magnitude of the recoil peak is smaller in agreement with experiments.

\begin{figure}
\begin{center}
\epsfxsize=8cm
\epsfbox{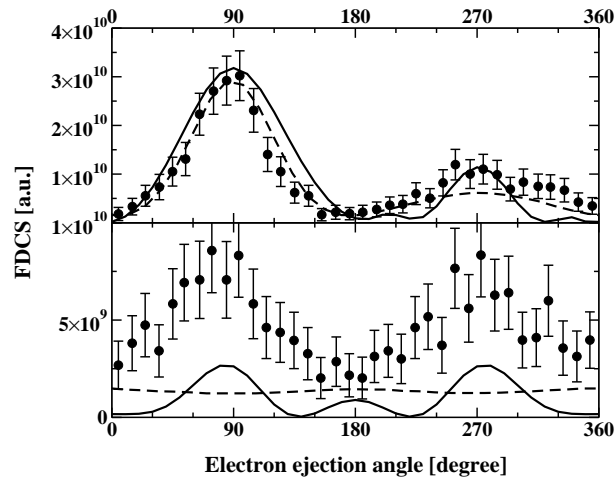}
\end{center}
\caption{\label{0.75planes}Theoretical results in scattering (top) and 
perpendicular (bottom) planes compared with experiments \cite{Fiol2006} for the 
same case as in figure \ref{3d1}. The solid curve shows the present theory 
while the dashed line is obtained by the CDW model \cite{Fiol2006}.}
\end{figure}

In order to analyze in detail the obtained results, cross section values for 
scattering plane and perpendicular plane are plotted separately in figure 
\ref{0.75planes}. The results obtained by the present theory may be compared in
absolute value to experimental data and CDW calculations. The top panel shows
fully differential cross sections in scattering plane. The semi-empirical model
gives good 
agreement with experimental data of Schulz et al. \cite{Schulz2003a}. In 
contrast to the previous calculations using a single value for the impact 
parameter, a smaller magnitude for the recoil peak has been obtained. However, 
we have to note that the shape of the binary peak is slightly wider than the 
experimental one.

Better results in case of perpendicular plane have also been obtained (bottom 
panel of figure \ref{0.75planes}). The curve shows the same behavior as the 
experimental data with strong maxima at $\theta = 80^o$ and $\theta = 280^o$. A 
third smaller maximum is also obtained at direction of $\theta = 180^o$. Here 
we have to note that better agreement in shape has been obtained than the 
isotrope results of the CDW model. However, the magnitude of the cross section 
is smaller than the experimental one in the perpendicular plane. The recently 
reported inclusion of the experimental momentum uncertainties \cite{Fiol2006} 
should also improve the agreement between theory and experiments by increasing 
the cross sections in the perpendicular plane. Our result is consistent with
the conclusions of D\"urr et al. \cite{Durr2007}, that the experimental
uncertainties are responsible only partly for the structure observed in the
perpendicular plane, half of the value of the maxima may be due to some real
physical effect.

\begin{figure}
\begin{center}
\epsfxsize=8cm
\epsfbox{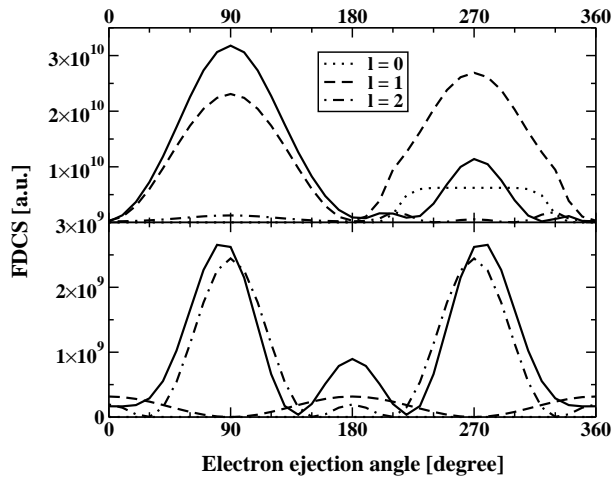}
\end{center}
\caption{\label{0.75detail}Multipole contributions to the ionization cross 
sections in scattering (top) and perpendicular (bottom) planes for ejected 
electron energy $E_e = 6.5$ eV and momentum transfer $q=0.75$ a.u. in 
comparison with the FDCS values of the present theory (solid line). Cross 
sections for different transition mechanisms are drawn separately (see text).}
\end{figure}

Another analysis has also been performed in order to clarify which type of 
transitions are responsible for the obtained structures. Cross sections 
corresponding to different terms of the multipole expansion series are shown in 
figure \ref{0.75detail}.

Let us first discuss the results in the scattering plane depicted in the top panel.
 The main contribution to the cross section (solid line) has the $l=1$ dipole 
term. Moreover, this term gives a large contribution in case of recoil peak, 
which is reduced by the destructive interferences with the monopole term 
($l=0$) which has contribution only in recoil peak region calculated with a 
smaller impact parameter. Terms with $l\geq2$ have negligible contribution to 
the fully differential cross section values in scattering plane.

In contrast, in case of perpendicular plane (bottom panel of figure 
\ref{0.75detail}) the main contributing term is the $l=2$ quadrupole term from 
the multipole expansion of the perturbation potential. This term is responsible 
for the shape of the electron emission pattern in this plane. The corresponding 
contribution has maxima at $\theta = 90^o$ and $\theta = 270^o$.
These 
maxima are shifted to 80$^o$ and 280$^o$ due to the interferences with other 
multipole terms. This shifting is also detectable in the experimental results. 
The monopole term practically has no contribution to the cross sections in this 
plane. A constructive interference occurs at $\theta = 180^o$ being responsible 
for the additional maximum occurring in the theoretical results. Terms with 
$l\geq3$ have negligible contribution to the fully differential cross section 
values in perpendicular plane, too.

\begin{figure}
\begin{center}
\epsfxsize=12cm
\epsfbox{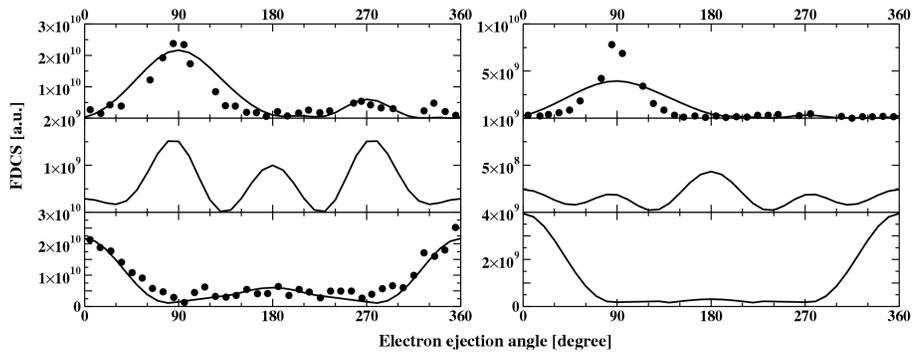}
\end{center}
\caption{\label{planes}Theoretical FDCS values in scattering plane (top 
graphs), in plane perpendicular to the momentum transfer (middle graphs) and in 
plane perpendicular to the beam direction in comparison with experiments 
\cite{Madison2002} for $E_e = 6.5$ eV and $q=0.88$ 
a.u. (left panel) and for $E_e = 17.5$ eV and $q=1.43$ a.u. (right panel).}
\end{figure}

The next studied case is with ejected electron energy $E_e = 6.5$ eV and 
momentum transfer of $q=0.88$ a.u., where the experimental results in 
scattering plane show cross sections of $2.36\times 10^{10}$ a.u and $4.2 
\times 10^{9}$ a.u. for binary peak and recoil peak, respectively. In 
calculations impact parameters of 1.7 and 0.6 a.u are used determined by the 
above described semi-empirical method. Cross section values for scattering and 
perpendicular plane are plotted separately in the left panel of the figure 
\ref{planes}. The top graphic shows fully differential cross sections in the 
scattering plane. The semi-empirical model gives good agreement with 
experimental data \cite{Madison2002}. In case of the perpendicular plane (middle 
graph) no experimental data was found. However, the structure is  similar to 
the previous case with two maxima and another smaller maximum at  $180^o$. It 
has to be mentioned that the difference between these two type of maxima is 
reduced. The bottom graph shows theoretical results in xy plane perpendicular 
to the incident beam. The results are in good agreement with the experiments.

The last studied case is with ejected electron energy $E_e = 17.5$ eV and 
momentum transfer of $q=1.43$ a.u., where the experimental results in 
scattering plane show cross sections of $7.34\times 10^{9}$ a.u and $3.71 
\times 10^{8}$ a.u. for binary peak and recoil peak regions. In theoretical 
calculations impact parameters of 0.8 and 0.4 a.u are chosen. Here some
discrepancies between theory and experiments can be found in scattering 
plane presented in right panel of the figure \ref{planes}. 
The theoretical curve for scattering plane is wider and smaller than the 
experimental data. In perpendicular plane the maximum at 180$^o$ 
became grater than for smaller momentum transfers, and another additional
maximum is appearing at 0$^o$. The model suggests that these maxima occurs
mainly due to the quadrupole transitions amplified by constructive
interferences. According to the paper \cite{Durr2007} the
experimental uncertainties have no important effect in the structures observed in
the experimental data. Our calculations predict in this perpendicular plane
similar structure to the experimental ones \cite{Durr2007} but we could not
compare them in the absolute scale, because the data are published for this
momentum transfer in arbitrary units.

In conclusion, a theoretical model based on the first order, semiclassical, 
impact parameter approximation has been constructed to simulate kinematically 
complete experiments and was applied for studying single ionization of helium 
by impact with fast C$^{6+}$ ions. A semiempirical model was developed, which 
uses large impact parameters for reproducing the binary peak and smaller impact 
parameters for the recoil peak. The model describes well the fully differential 
cross sections for relatively small momentum transfer values. The 
characteristic structures in perpendicular plane have also been reproduced, 
discrepancies with experiments are only in the magnitude of the cross sections. 
The other part of the cross section values may be explained by the experimental 
uncertainties. It was found that in scattering plane the main contribution has the 
dipole transition term, while  in perpendicular plane the characteristic 
structure is mainly due to the quadrupole transitions. Another important 
observation is that interferences between the multipole expansion terms are 
also important to understand the exact structure of the electron emission 
patterns. Our semiclassical model includes projectile--nucleus scattering, and
we may conclude that this should be important in obtaining the experimentally
observed FDCS structures in the perpendicular plane.

The authors acknowledge the support of the Romanian National Plan for Research (PN II) under contract No ID\_539.

\end{document}